\documentclass[a4paper,UKenglish,cleveref, autoref, thm-restate]{lipics-v2021}
\hideLIPIcs  %uncomment to remove references to LIPIcs series (logo, DOI, ...), e.g. when preparing a pre-final version to be uploaded to arXiv or another public repository

\usepackage{amsmath,amssymb,amsfonts}
\usepackage{algorithm}
\usepackage{algorithmic}
\usepackage{graphicx}
\usepackage{svg}
\usepackage{textcomp}
\usepackage{xcolor}
\usepackage{tcolorbox}
\usepackage{hyperref}

\usepackage{braket} % added by miao (2021/12/12)

\usepackage{booktabs} % For formal tables
\usepackage{multirow}
\usepackage{float}
\newfloat{Figure}{htbp}{lof}[section]
\usepackage{diagbox}

\usepackage{url}
\usepackage{stmaryrd}
\definecolor{deepred}{RGB}{155,40,40}
\usepackage{listings}
\lstdefinestyle{mystyle}{
    moredelim=**[is][\color{gray}]{|}{|}, 
    moredelim=**[is][\color{deepred}]{^}{^}, 
    moredelim=**[is][\color{blue}]{~}{~}, 
}
\usepackage{cleveref}
\usepackage{amsfonts}
\usepackage{makecell}

\usepackage{tikz}
\usetikzlibrary{quantikz2}

\usepackage{graphicx}
\usepackage{epsf}
\usepackage{verbatim}
\usepackage{url}
\usepackage{color}
\usepackage{alltt}

\newcommand{\Comment}[1]{}

\newcommand{\SmallSpace}{\vspace*{-1.4ex}}

\title{{Dataflow-Based Optimization for Quantum Intermediate Representation Programs}} %TODO Please add

\author{Junjie Luo}{Kyushu University, Japan}{}{}{}%{(Optional) author-specific funding acknowledgements}

\author{Haoyu Zhang}{Kyushu University, Japan}{}{}{}%{(Optional) author-specific funding acknowledgements}

\author{Jianjun Zhao}{Kyushu University, Japan}{}{https://orcid.org/0000-0001-8083-4352}{}%{(Optional) author-specific funding acknowledgements}%TODO mandatory, please use full name; only 1 author per \author macro; first two parameters are mandatory, other parameters can be empty. Please provide at least the name of the affiliation and the country. The full address is optional. Use additional curly braces to indicate the correct name splitting when the last name consists of multiple name parts.

\authorrunning{ } %TODO mandatory. First: Use abbreviated first/middle names. Second (only in severe cases): Use first author plus 'et al.'

 %TODO mandatory: Please choose ACM 2012 classifications from https://dl.acm.org/ccs/ccs_flat.cfm 

 %TODO mandatory; please add comma-separated list of keywords

\category{} %optional, e.g. invited paper

\relatedversion{} %optional, e.g. full version hosted on arXiv, HAL, or other respository/website
\nolinenumbers %uncomment to disable line numbering

\begin{document}

\maketitle

\begin{abstract}
This paper proposes QDFO, a dataflow-based optimization approach to Microsoft QIR. QDFO consists of two main functions: one is to preprocess the QIR code so that the LLVM optimizer can capture more optimization opportunities, and the other is to optimize the QIR code so that duplicate loading and constructing of qubits and qubit arrays can be avoided. We evaluated our work on the IBM Challenge Dataset, the results show that our method effectively reduces redundant operations in the QIR code. We also completed a preliminary implementation of QDFO and conducted a case study on the real-world code. Our observational study indicates that the LLVM optimizer can further optimize the QIR code preprocessed by our algorithm. Both the experiments and the case study demonstrate the effectiveness of our approach. 
\end{abstract}

\section{Introduction}
\textcolor{black}{In recent years, the rapid advancement of quantum computing technology has sparked widespread interest in quantum programming and quantum program compilation~\cite{cross2017open,Green_2013,mccaskey2019xacc,Svore_2018,8715261}. Traditional programming paradigms and compilation techniques, when applied directly to quantum computing, encounter significant challenges due to the inherent properties of quantum bits (qubits for short), such as superposition and entanglement, which introduce complexities far beyond those encountered in classical computing~\cite{10.1145/3579367,Zalka_1998,Montanaro_2016}. Researchers have developed specialized programming models, compilation techniques, and software toolchains tailored specifically for quantum computing to harness the potential of quantum computation~\cite{garhwal2019quantum,zorzi2019quantum,fingerhuth2018open}. }

\textcolor{black}{As quantum hardware continues to improve, the quantum programming environment is concurrently evolving and maturing, encompassing various components of the quantum development ecosystem, including quantum programming languages (such as \texttt{Q\#}~\cite{Svore_2018} and \texttt{Quipper}~\cite{Green_2013}), quantum development kits (like \texttt{Qiskit}~\cite{8715261} and \texttt{Azure QDK}~\cite{Azure_QDK}), and quantum compilation frameworks (such as \texttt{XACC}~\cite{mccaskey2019xacc}). Recently, the introduction of Quantum Intermediate Representation (QIR) by Microsoft~\cite{alan_2020_qir} marks a significant step forward in quantum programming. QIR serves as a bridge between quantum programming languages and the underlying quantum hardware, providing developers with a streamlined interface that simplifies the creation and optimization of quantum programs across various quantum computing platforms.}

\textcolor{black}{In this paper, we delve into the effectiveness of optimization algorithms that target QIR in quantum program compilation. In the era of Noisy Intermediate-Scale Quantum (NISQ) computing~\cite{Preskill_2018}, where devices have limited qubit coherence times and high error rates, the challenges of compilation and optimization become more pronounced. Therefore, developing optimization algorithms that address these unique challenges is crucial for unlocking the potential of practical quantum computing applications. 
To this end, we propose a novel dataflow-based optimization approach called Quantum Intermediate Representation DataFlow Optimization (QDFO) built upon the LLVM infrastructure. By seamlessly integrating QDFO into the LLVM compiler framework, we aim to tackle the unique challenges of optimizing QIR programs.}
\textcolor{black}{Our main contributions are outlined as follows:}

\begin{itemize}%[leftmargin=1.5em]
\setlength{\itemsep}{2pt}
\item \textcolor{black}{\textbf{Dataflow-based optimization algorithms for QIR code}: We developed and implemented the Quantum Intermediate Representation Data Flow-based Optimization Algorithm (QDFO), leveraging the LLVM infrastructure to optimize the QIR programs based on dataflow. The algorithms enhance the efficiency and performance of quantum computations.}

\item \textcolor{black}{\textbf{Integration of our algorithms with LLVM infrastructure}: We successfully integrated the QDFO algorithm into the LLVM compiler framework, facilitating seamless interaction with existing LLVM optimization passes and toolchains. This integration significantly enhances the accessibility and usability of QDFO within the broader LLVM ecosystem.}

\item \textcolor{black}{\textbf{Case studies on real-world examples}: We researched the QDFO algorithm using two real-world quantum programs. Through our observations, we can confirm the effectiveness and efficiency of QDFO in optimizing QIR programs, and it also improves the readability of the QIR code.}
\end{itemize}

%\textcolor{black}{By presenting these contributions, we aim to contribute to the ongoing discussion in quantum computing and provide valuable insights for researchers and practitioners.}

The rest of the paper is organized as follows. We provide a brief background of our work in Section \ref{sec:BG} and present a motivating example in Section \ref{sec:ME}. The mechanism of QDFO is elucidated in Section \ref{sec:methodology}. \textcolor{black}{We conducted case studies of the real-world QIR code in Section~\ref{sec:EE}}, review related work in Section \ref{sec:RW}, and summarize our work in Section \ref{sec:CR}.

%Our work is publicly available at 
%\url{https://anonymous.4open.science/r/QIR_PASS-576C}

\section{Background}\label{sec:BG}
This section will briefly introduce the concepts and foundational knowledge of Intermediate Representation (IR), LLVM, Quantum Intermediate Representation (QIR), and Quantum Computing, aiming to facilitate comprehension for subsequent sections.

\subsection{Intermediate Representation}
An intermediate representation (IR) denotes the data structure or coding paradigm utilized internally within a compiler or virtual machine to represent source code. IR is meticulously crafted to facilitate subsequent operations, including optimization and translation.

In compiler design, a prevalent strategy involves translating the source language into an IR. This intermediary format is carefully crafted to accommodate diverse source languages. Within this intermediate phase, the code undergoes optimization and metamorphosis. Subsequently, upon identification of the definitive target execution platform, the intermediate representation is transmuted into tangible executable code.

Such an approach fosters utilizing a shared set of optimizers and executable generators across numerous source languages. Moreover, it streamlines the process of compiling a singular source language for manifold targets. The intermediate representation facilitates extensive reusability within the compiler infrastructure by furnishing a unified platform spanning multiple sources and targets.

\subsection{LLVM}
The LLVM framework~\cite{1281665} represents a landmark achievement in compiler technology, renowned for its robust infrastructure and versatility. At its core, LLVM embodies a modular design philosophy, comprising various components such as the LLVM Intermediate Representation (IR), a versatile set of compiler tools, and a powerful optimization infrastructure.

\subsubsection{LLVM IR} Central to the LLVM framework is its Intermediate Representation (IR), a language-agnostic, low-level code representation. Unlike traditional compilers that operate directly on source or assembly code, LLVM operates on its IR, enabling various analyses and transformations irrespective of the source language.

In LLVM, the Def-Use (DU) and Use-Def (UD) chains are fundamental mechanisms for analyzing and optimizing programs' IR.

The Def-Use chain represents the relationship between definitions (where a variable or value is set) and their uses (where the variable or value is read). It essentially tracks how the values flow from their definitions to their uses within the program. This chain is typically used for various analyses and optimizations, such as dead code elimination, common subexpression elimination, and register allocation.

On the other hand, the Use-Def chain represents the reverse relationship, from uses back to definitions. It provides the ability to quickly locate the definition(s) of a value used at a particular point in the program. This chain is handy in optimizations like constant propagation and value numbering.

\subsubsection{LLVM Optimizer} LLVM Optimizer is a crucial component of the LLVM compiler infrastructure, designed to enhance the performance of code generated by LLVM. The optimizer applies a series of transformations to the IR to improve code efficiency, reduce execution time, and minimize resource usage.

Within LLVM Optimizer, passes are the fundamental units of optimization. Each pass performs a specific transformation on the IR, targeting various aspects of code quality and performance. These passes encompass a wide range of optimizations, including but not limited to instruction simplification, control flow analysis, dataflow analysis, loop transformations, and target-specific optimizations. Notably, LLVM provides a flexible framework for customizing passes, allowing developers to tailor optimizations to specific needs and target architectures. This customization enables fine-tuning optimization strategies to address particular code patterns or performance bottlenecks. Taking advantage of the versatility of passes, the QDFO algorithm is implemented through the customization of pass modules in LLVM.

LLVM optimizers typically optimize for functions, reflecting the modular nature of the LLVM optimization pipeline. This approach allows optimizations to be applied locally within individual functions, facilitating targeted improvements without affecting the overall program structure. By focusing on function-level optimization, LLVM can balance optimization effectiveness and compilation efficiency, ensuring that optimizations contribute meaningfully to overall code performance while maintaining reasonable compilation times.

\subsection{Quantum Intermediate Representation}

Quantum Intermediate Representation (QIR)~\cite{alan_2020_qir}, an emerging intermediate representation devised by Microsoft for quantum programming, draws its foundations from the widely adopted LLVM intermediate language. This innovative framework delineates a set of regulations for encapsulating quantum constructs within LLVM, sans necessitating any extensions or alterations to the LLVM architecture. 

Primarily designed to function as a universal interface across diverse programming languages and quantum computation platforms, QIR transcends its association with \texttt{Q\#} to accommodate any language tailored for gate-based quantum computing. Furthermore, QIR maintains a hardware-agnostic stance by refraining from dictating a specific quantum instruction or gate set, thereby deferring such specifications to the discretion of the target computing environment.

With the maturation of quantum computing capabilities, the paradigm shift towards leveraging both classical and quantum resources is anticipated for most large-scale quantum applications. By leveraging LLVM, QIR seamlessly integrates rich classical computation with quantum counterparts, thereby availing comprehensive computational capabilities. The utilization of LLVM also fosters seamless integration with an array of classical languages and tools already supported by the LLVM toolchain. Moreover, it promotes the cultivation of standard, language-agnostic optimizations, and code transformations underpinned by a well-established and robust open-source framework.

\section{Motivating Example}
\label{sec:ME}

%\textcolor{red}{show how the Q\# code transfer to the QIR code?}

\begin{figure}[!htp]
    \centering
\begin{quantikz}
qs[0]\ &&&&&\ctrl{2}&&&&\ctrl{2}\gategroup[3,steps=5,style={dashed,rounded corners,fill=red!20,inner xsep=2pt},background,label style={label position=below,anchor=north,yshift=-0.2cm}]{{\sc Figure ~\ref{fig:storecode}}}&&\ctrl{1}&\gate{T}&\ctrl{1}& \\ 
qs[1]\ &&\gategroup[2,steps=2,style={dashed,rounded corners,fill=blue!20,inner xsep=2pt},background,label style={label position=below,anchor=north,yshift=-0.2cm}]{{\sc Figure ~\ref{fig:loadcode}}}&\ctrl{1}&&&&\ctrl{1}&&&\gate{T}&\targ{}&\gate{T^\dagger}&\targ{}& \\
qs[2]\ &&\gate{H}&\targ{}&\gate{T^\dagger}&\targ{}&\gate{T}&\targ{}&\gate{T^\dagger}&\targ{}&\gate{T}&\gate{H}&&&
\end{quantikz}
    \caption{A decomposition of a Toffoli gate. In the QIR code for this circuit, 21 operations to load qubits from a qubit array and six operations to construct control qubit arrays are performed. In contrast, ideally, only 3 and 2 operations are required.}
    \label{fig:toffoligate}
\end{figure}
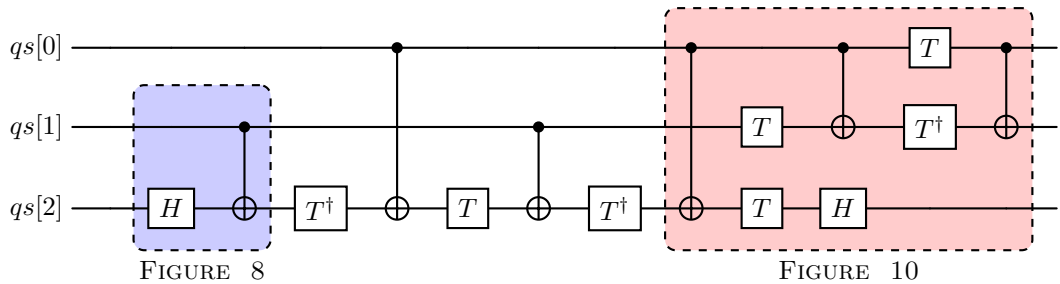

\begin{figure}[!htb]
\centering
\begin{lstlisting}[caption=QIR code before O3 optimization, basicstyle=\scriptsize\ttfamily,  numbers=left, style=mystyle, 
%linerange={22,28-30,34,119-131}, 
xleftmargin=-0.5em, 
xrightmargin=-3em, 
label={lst:beforeO3}, 
breaklines=true]
...
  |%qs = call %Array* @qubit_allocate_array(i64 3)|
  |%2 = call i8* @array_get_element_ptr_1d(%Array* %qs, i64 1)|
  |%3 = bitcast i8* %2 to %Qubit**|
  |%4 = load %Qubit*, %Qubit** %3, align 8|
  |%5 = tail call i8* @array_get_element_ptr_1d(%Array* %qs, i64 2)|
  |%6 = bitcast i8* %5 to %Qubit**|
  |%7 = load %Qubit*, %Qubit** %6, align 8|
  ^call void @Intrinsic__CNOT__body(%Qubit* %4, %Qubit* %7)^
...
  
define internal void @Intrinsic__CNOT__body(%Qubit* %control, %Qubit* %target) {
entry:
  %__controlQubits__ = call %Array* @array_create_1d(i32 8, i64 1)
  %0 = call i8* @array_get_element_ptr_1d(%Array* %__controlQubits__, i64 0)
  %1 = bitcast i8* %0 to %Qubit**
  store %Qubit* %control, %Qubit** %1, align 8
  |call void @array_update_alias_count(%Array* %__controlQubits__, i32 1)|
  call void @qis__x__ctl(%Array* %__controlQubits__, %Qubit* %target)
  |call void @array_update_alias_count(%Array* %__controlQubits__, i32 -1)|
  |call void @array_update_reference_count(%Array* %__controlQubits__, i32 -1)|
  ret void
}

\end{lstlisting}

\begin{lstlisting}[caption=QIR code after O3 optimization, basicstyle=\scriptsize\ttfamily,  numbers=left, style=mystyle, 
%linerange={22,28-30,34,119-131}, 
xleftmargin=-0.5em, 
xrightmargin=-3em, 
label={lst:afterO3}, 
breaklines=true]
...
  |%qs = call %Array* @qubit_allocate_array(i64 3)|
  |%2 = call i8* @array_get_element_ptr_1d(%Array* %qs, i64 1)|
  |%3 = bitcast i8* %2 to %Qubit**|
  |%4 = load %Qubit*, %Qubit** %3, align 8|
  |%5 = tail call i8* @array_get_element_ptr_1d(%Array* %qs, i64 2)|
  |%6 = bitcast i8* %5 to %Qubit**|
  |%7 = load %Qubit*, %Qubit** %6, align 8|
  ^%__controlQubits__.i = tail call %Array* @array_create_1d(i32 8, i64 1)^
  ^%8 = tail call i8* @array_get_element_ptr_1d(%Array* %__controlQubits__.i, i64 0)^
  ^%9 = bitcast i8* %8 to %Qubit**
  store %Qubit* %4, %Qubit** %9, align 8^
  |tail call void @array_update_alias_count(%Array* %__controlQubits__.i, i32 1)|
  ^tail call void @qis__x__ctl(%Array* %__controlQubits__.i, %Qubit* %7)^
  |tail call void @array_update_alias_count(%Array* %__controlQubits__.i, i32 -1)|
  |tail call void @array_update_reference_count(%Array* %__controlQubits__.i, i32 -1)|
...


\end{lstlisting}
    \caption{The QIR code before and after a process of LLVM O3 optimization. }
    \centering
    \label{fig:O3optimization}
\end{figure}

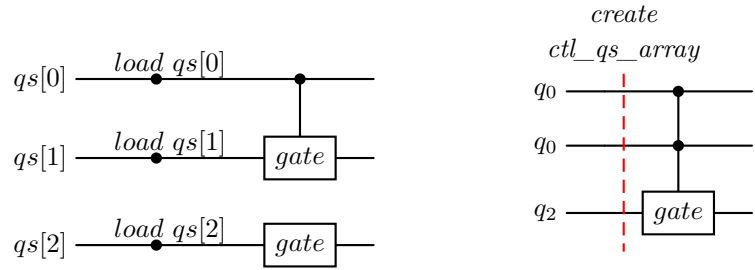
\begin{figure}[htb]
    \centering
\begin{lstlisting}[caption=Instructions required to load a qubit from a qubit array in QIR ($Load\_Op$), basicstyle=\scriptsize\ttfamily,  numbers=left, style=mystyle, 
%linerange={22,28-30,34,119-131}, 
%xleftmargin=-0.5em, 
%xrightmargin=-3em, 
label={lst:loadsample}, 
breaklines=true]
  |%qs = tail call %Array* @qubit_allocate_array(i64 3)|
  %0 = tail call i8* @array_get_element_ptr_1d(%Array* %qs, i64 2)
  %1 = bitcast i8* %0 to %Qubit**
  %qubit = load %Qubit*, %Qubit** %1, align 8

\end{lstlisting}
\begin{lstlisting}[caption=Instructions required to create a qubit array from allocated qubits in QIR ($Create\_Op$), basicstyle=\scriptsize\ttfamily,  numbers=left, style=mystyle, 
%linerange={22,28-30,34,119-131}, 
%xleftmargin=-0.5em, 
%xrightmargin=-3em, 
label={lst:storesample}, 
breaklines=true]
  %__controlQubits__ = tail call %Array* @array_create_1d(i32 8, i64 1)
  %0 = tail call i8* @array_get_element_ptr_1d(%Array* %__controlQubits__, i64 0)
  %1 = bitcast i8* %0 to %Qubit**
  store %Qubit* %qubit, %Qubit** %1, align 8

\end{lstlisting}
    \begin{quantikz}
        qs[0]\ &\phase{load\ qs[0]}\ &\ctrl{1}&\\
        qs[1]\ &\phase{load\ qs[1]}\ &\gate{gate}& \\
        qs[2]\ &\phase{load\ qs[2]}\ &\gate{gate}&
    \end{quantikz}
    \quad\quad\quad\quad\quad
    \begin{quantikz}
        q_0\ &\slice{\textit{create}\\ \textit{ctl\_qs\_array}}&\ctrl{1}&\\
        q_0\ &&\ctrl{1}&\\
        q_2\ &&\gate{gate}& 
    \end{quantikz}
    \caption{In QIR, if the qubits for a quantum gate operation come from a qubit array, the instructions to load qubits from the qubit array need to be executed one time on these qubits before each execution of the quantum gate operation (lines 2-4 in Listing~\ref{lst:loadsample}). For each controlled quantum gate, the instruction to generate the corresponding control qubit array needs to be executed before performing the control quantum gate operation (Listing~\ref{lst:storesample}).}
    \label{fig:loadstoresample}
\end{figure}

We present the demands for optimization of QIR codes through a decomposition of the Toffoli gate~\cite{Shende2008OnTCToffolidecompose}. Figure~\ref{fig:toffoligate} illustrates the decomposed circuit, which achieves the same functionality as compared to the original Toffoli gate, with the difference being that only single and two-qubit gates are used in the construction of the decomposed circuit, resulting in lower requirements for the hardware on which it is operating. 

In this example, we implemented the circuit with the \texttt{Q\#} programming language and generated the corresponding QIR code with the compiler provided by \texttt{Q\#}. Due to the length of the QIR code, we will only show the code for the first CNOT gate in Figure~\ref{fig:toffoligate} as a complete presentation in Listing~\ref{lst:beforeO3}. Specifically, it first allocates a qubit array of length 3 (line 2), then loads the 2nd and 3rd qubits in the array (lines 3-5, lines 6-8), respectively, and passes them to a CNOT gate function (line 9). In the implementation of the CNOT gate function (lines 12-23), it first creates an empty array of length 1 (line 14), stores the pointer to the controlling qubit in that array (lines 15-17), and executes the QIR CNOT gate instruction (line 19). The remainder of the function is the memory management of the created array (line 18, lines 20-21). 

Here, we are mainly concerned with the QIR code's management of qubit arrays, which consists of loading a single qubit from an allocated qubit array and constructing a new quantum array with the existing qubits. We summarize these two operations in Figure~\ref{fig:loadstoresample} and refer to them as $Load\_Op$ and $Create\_Op$. Intuitively, for the circuit of Figure~\ref{fig:toffoligate}, we only need to load the qubits in the array once for each of them to obtain their corresponding register addresses. Then, we can use these addresses repeatedly when executing the gate operation without loading the same qubits repeatedly. However, in the actual QIR code, such loading behavior will be executed once every time a gate operation is used because the compiler does not remember which qubits in the hardware device each qubit register address is but instead obtains the corresponding qubit register addresses from the quantum array before every gate operation is executed. It means that $Load\_Op$ needs to be performed either once or twice for each single-qubit gate and two-qubit gate, resulting in 21 times $Load\_Op$ being conducted in the QIR code generated in Figure~\ref{fig:toffoligate}.
Similarly, even though a total of only two control qubit arrays (containing qs[0] and qs[1], respectively) appear in the circuit of Figure~\ref{fig:toffoligate}, a total of 6 control qubit arrays are created in the actual QIR code, which means that $Create\_Op$ is executed six times. Since $Load\_Op$ and $Create\_Op$ occur along with gate operations, the operations for such redundancy increase linearly as the circuit size increases. Consequently, this affects the size and correctness of the quantum program that can be run on the current NISQ hardware. 

As an initial validation of our optimization approach, we first performed an O3-level optimization of this QIR code as a preprocessor using the optimizer in the LLVM toolchain and then optimized it using our optimization approach. Intuitively, the resulting QIR code is significantly reduced in code size, with its implementation of the Toffoli gate decomposition circuit reduced from 128 lines to 46 lines after optimization specifically, where $Load\_Op$ and $Create\_Op$ decreased from 21 and 6 times to 3 and 2 times. At the same time, the program's behavior did not change.

\section{Methodology}
\label{sec:methodology}

%\textcolor{red}{modify the description of the algorithm}

We can find optimization chances for current QIR codes by exploring the cases in Section~\ref{sec:ME}. When dealing with classic LLVM IR code, we apply optimization methods such as function inlining and loop unrolling to the code, thus providing more contextual information for other optimization methods and optimization opportunities. Such experience can also be transferred to the processing of QIR code; for example, the content of the instruction $@Intrinsic\_\_CNOT\_\_ctl$ can be directly replaced at the function call through function inlining so that we can determine whether the same control qubit array is generated repeatedly in the program, and then further optimization can be carried out.

Since QIR is developed based on LLVM IR without any modifications or extensions to LLVM, LLVM's optimization methods can also be directly applied to QIR's code to optimize it. However, using only LLVM's built-in optimization methods only captures a few opportunities to optimize QIR code. This is because, similar to the design idea of the \texttt{Q\#} programming language, QIR treats the Qubit as an opaque object and operates on the actual qubit by passing only the address of the Qubit register to the instruction of the gate operation~\cite{qsharpqubits}. Specifically, the state change of the actual qubits is controlled by the side-effect of the instruction, and it is impossible to explicitly change the value of the qubits directly through the instruction. The use of this operation avoids entanglement confusion~\cite{entanglementconfusion} while also posing challenges to the LLVM's optimizer: the first is that the LLVM's optimizer is unable to directly obtain the contents of opaque values in the QIR, e.g., the length of a \texttt{\%Array} can only be obtained through the instruction $@array\_get\_size\_1d$ during dynamic execution, which makes some optimization methods like loop unrolling and function inlining limited during static optimization. Secondly, to satisfy QIR's hardware-agnostic, QIR-specific instructions are left to be implemented by the target computing environment and called by \texttt{call} instructions in QIR code in the form of functions. It is reasonable for quantum instructions and gate operations, which control hardware qubits by the side effect and, therefore, avoid the problem of their accidental modification or deletion by the LLVM optimizer due to the lack of explicit return values. However, some instructions like $@get\_element\_ptr\_1d$ and $@array\_get\_size\_1d$, which only return values and do not operate on objects, can lead to the problem of not being optimized by the DCE (dead code elimination) algorithm~\cite{DCE} even if the value obtained by the instruction is no longer used after the optimization. 

Therefore, in this section, we first introduce the workflow of our optimizations (Section~\ref{subsec:workflow}) and a brief overview of the application of LLVM optimization to QIR (Section~\ref{subsecLLVMOp}). Then we design a series of optimization methods for QIR, which include additions to the LLVM optimization methods in scenarios where QIR code is processed (Sections~\ref{subsec:QIRInline} and~\ref{subsec:QIRLU}), as well as new optimization methods based on dataflow (Section~\ref{subsec:QDFO}).

\subsection{QDFO Workflow}
\label{subsec:workflow}

\begin{figure}[htb]
    \centering
    \includegraphics[width=1.0\textwidth]{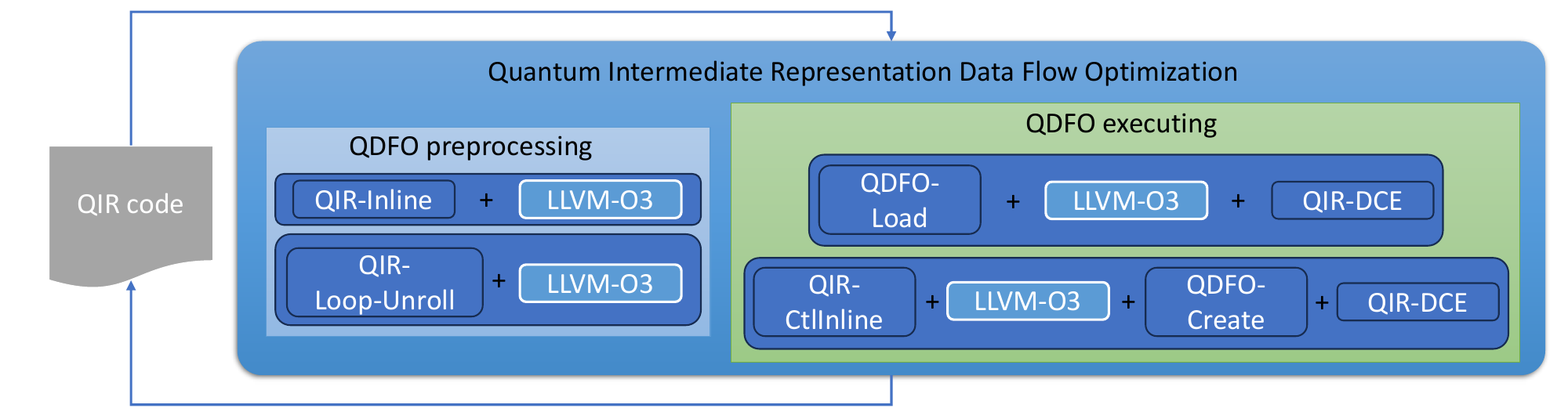}
    \caption{The workflow of Quantum Intermediate Representation Data Flow-based Optimization. }
    \label{fig:workflow}
\end{figure}

Our work consists of two main modules (see Figure~\ref{fig:workflow}):
\begin{itemize}
    \item \textbf{QDFO preprocessing} includes two main processing steps, \textbf{QIR-Inline} and \textbf{QIR-Loop-Unroll}. This two-step operation statically analyzes the code by replacing the QIR-specific function calls and loop control instructions that need to be obtained through QIR's instructions with function calls and constants that are common to LLVM IR so that regular function inlining and loop unrolling optimizations can be performed with LLVM's O3-level optimization.
    \item \textbf{QDFO executing} contains two main processing steps, \textbf{QDFO-Load} and \textbf{QDFO-Create}. The \textbf{QDFO-Load} optimization algorithm can change the use of duplicate $Load\_Op$ in the code to the qubit register address obtained from the first $Load\_Op$, thus making these duplicate $Load\_Op$ into dead code. After performing the \textbf{QDFO-Load} optimization, an LLVM's O3-level optimization and a \textbf{QIR-DCE} optimization need to be performed to remove the leftover dead code. The \textbf{QDFO-Create} optimization algorithm, on the other hand, can change the use of duplicate $Create\_Op$'s in the code to all the address of the qubit array obtained from the first $Store\_Op$ and remove the latter's duplicate $Create\_Op$'s as dead code. Before performing the \textbf{QDFO-Create} optimization, it is necessary to give an \texttt{always-inline} attribution (\textbf{QIR-CtlInline}) to the functions of the control gates and SWAP gates in the code since the control gates contain both the $Load\_Op$ and $Create\_Op$ operations, which have a higher number of optimization opportunities. In contrast, the SWAP is implemented in the QIR code through the three CNOT gate operations and can be optimized. After that, an LLVM-O3 optimization must be performed to complete the function inlining.
\end{itemize}

When executing our optimization algorithm, it is only necessary to repeatedly apply it to the QIR code until the size of the code no longer changes. Then, the optimization process can be stopped. From our practical experience, we find that the maximum optimization effect can be achieved by executing the algorithm at most twice because there are fewer complex nested loops or nested functions in the current QIR program.

\subsection{LLVM Optimizer Optimization}
\label{subsecLLVMOp}

LLVM's optimizer has been developed over a long period since its inception. Its stability and robustness are guaranteed, and it can also apply multiple optimization techniques at different levels. For this reason, we actively utilize the optimization methods that come with the LLVM optimizer, such as function inlining and loop unrolling. Although applying these methods directly to QIR code has little effect, their potential for optimization on QIR code can be exploited as much as possible through the processing of our algorithm. 

Performing function inlining and loop unrolling for QIR code is beneficial for optimizing QIR programs. Figure~\ref{fig:O3optimization} shows a typical example where we place the function body of the function $@Intrinsic\_\_CNOT\_\_body$ at the location where it is called (from Listing~\ref{lst:beforeO3} to Listing~\ref{lst:afterO3}) by inlining the function. We can obtain information about $\%\_\_controlQubits\_\_.i$, used by the CNOT gate instruction in line 14, thus providing an opportunity for subsequent optimization.
%We will describe in the following sections how our approach makes changes to QIR code so that the optimization methods of the LLVM optimizer can be applied.

\subsection{QDFO Preprocessing}
In this subsection, we introduce two optimizations of QIR code, \textbf{QIR-Inline} and \textbf{QIR-Loop-Unroll}, which allow the LLVM optimizer's optimization methods to be applied to QIR code, exposing more opportunities for optimizing the QIR programs.

\subsubsection{QIR-Inline}
\label{subsec:QIRInline}
%我们的工作只提供了一个循环展开和函数内联的机会，而实际上执行循环展开和函数内联的算法是llvm的O3优化算法，也就是说我们只需要论证值的替换和函数的替换是否合理，而无需证明循环展开和函数内联本身是否合理。

\begin{figure}[!htb]
\centering
\begin{lstlisting}[caption=QIR code before \textbf{QIR-Inline} optimization, basicstyle=\scriptsize\ttfamily,  numbers=left, style=mystyle, 
%linerange={22,28-30,34,119-131}, 
xleftmargin=-0.5em, 
xrightmargin=-3em, 
label={lst:beforeinline}, 
breaklines=true]
@something__FunctionTable = internal constant [4 x void (%Tuple*, %Tuple*, %Tuple*)*] [void (%Tuple*, %Tuple*, %Tuple*)* @something__body__wrapper, void (%Tuple*, %Tuple*, %Tuple*)* null, void (%Tuple*, %Tuple*, %Tuple*)* null, void (%Tuple*, %Tuple*, %Tuple*)* null]
%0 = tail call %Callable* @callable_create([4 x void (%Tuple*, %Tuple*, %Tuple*)*]* nonnull @something__FunctionTable, [2 x void (%Tuple*, i32)*]* null, %Tuple* null)
|%1 = tail call %Tuple* @tuple_create(i64 8)|
|%2 = bitcast %Tuple* %11 to %Array**|
|store %Array* %qubits.i, %Array** %2, align 8|
^tail call void @callable_invoke(%Callable* %0, %Tuple* %1, %Tuple* null)^

define internal void @something__body__wrapper(%Tuple* %capture-tuple, %Tuple* %arg-tuple, %Tuple* %result-tuple) { ... }

\end{lstlisting}

\begin{lstlisting}[caption=QIR code after \textbf{QIR-Inline} optimization, basicstyle=\scriptsize\ttfamily,  numbers=left, style=mystyle, 
%linerange={22,28-30,34,119-131}, 
xleftmargin=-0.5em, 
xrightmargin=-3em, 
label={lst:afterinline}, 
breaklines=true]
@something__FunctionTable = internal constant [4 x void (%Tuple*, %Tuple*, %Tuple*)*] [void (%Tuple*, %Tuple*, %Tuple*)* @something__body__wrapper, void (%Tuple*, %Tuple*, %Tuple*)* null, void (%Tuple*, %Tuple*, %Tuple*)* null, void (%Tuple*, %Tuple*, %Tuple*)* null]
%0 = tail call %Callable* @callable_create([4 x void (%Tuple*, %Tuple*, %Tuple*)*]* nonnull @something__FunctionTable, [2 x void (%Tuple*, i32)*]* null, %Tuple* null)
|%1 = tail call %Tuple* @tuple_create(i64 8)|
|%2 = bitcast %Tuple* %11 to %Array**|
|store %Array* %qubits.i, %Array** %2, align 8|
^call void @something__body__wrapper(%Tuple* %null, %Tuple* %1, %Tuple* null)^

define internal void @something__body__wrapper(%Tuple* %capture-tuple, %Tuple* %arg-tuple, %Tuple* %result-tuple) { ... }

\end{lstlisting}

    \caption{After a \textbf{QIR-Inline} process, function calls that would otherwise be QIR-specific become LLVM IR calls, allowing the execution of function inline optimization.}
    \centering
    \label{fig:QIRinline}
\end{figure}

Functions are usually called in LLVM IR using the \texttt{call} instruction. In QIR, an opaque \texttt{\%Callable} type has been added due to its generics support, which can be used for lambda captures and partial application, and a new wrapper function is created for each callable object specifically. The wrapper function is stored in a global function table, which is uniformly suffixed with "\_\_wrapper" and exists in four versions, which are: 
\begin{itemize}
    \item \textbf{body\_\_wrapper:} the original version of the function.
    \item \textbf{ctl\_\_wrapper:} the function that implements the controlled version of the quantum operation.
    \item \textbf{adj\_\_wrapper:} the function that implements the adjoint of the quantum operation.
    \item \textbf{ctladj\_\_wrapper:} the function that implements the adjoint of the controlled version of the quantum operation.
\end{itemize}

We use the \texttt{call} to body\_\_wrapper function in Figure~\ref{fig:QIRinline} as an example to introduce our approach to inline optimization for QIR. Here, we omit the code related to the memory management of the \texttt{\%Callable*} object generated in line 2 of the code, which will be described in detail in Subsections~\ref{par:QDCE} and~\ref{par:QDFOCreate}. 

From the description of $@callable\_invoke$ in QIR's specification, we know that when QIR tries to call the body\_\_wrapper function, it first needs to create a \texttt{\%Callable*} object (line 2 in Listing~\ref{lst:beforeinline}) and then use the $@callable\_invoke$ function to invokes body\_\_wrapper function (line 6 in Listing~\ref{lst:beforeinline}). The LLVM optimizer cannot inline calls by this method, but based on the semantics, we can directly replace the directive with a \texttt{call} to the body\_\_wrapper function at the point of use of the $@callable\_invoke$ function, where the required arguments to the wrapper function can be taken from the $@callable\_create$ and $@callable\_invoke$ functions. The modified QIR code is shown in Listing~\ref{lst:afterinline}.

\subsubsection{QIR-Loop-Unroll}
\label{subsec:QIRLU}

\textbf{QIR-Loop-Unroll} is dedicated to solving the problem where code tries to use a QIR-specific instruction to get the length of a qubit array and use it as a loop control statement, resulting in LLVM's loop unroll not working correctly. For example, the code in Listing~\ref{lst:beforeul} implements a function that ends the loop when $\%.not$ is less than 0 (line 6). By analyzing the use-def chain, it can be found that the value of $\%.not$ comes from $\%0$, which is the return value of $@array\_get\_size\_1d$'s processing of $\%qubits.i$ (line 3). As discussed earlier in this section, the LLVM optimizer cannot capture the return value of $@array\_get\_size\_1d$, so the loop cannot be analyzed for loop unrolling. And from the semantics of $@array\_get\_size\_1d$ we know that the return value of this function is the array length of qubit array $\%qubits.i$. So if we find the statement allocating $\%qubits.i$ with the help of the use-def chain (line 2), it is possible to get the exact length of the allocation, e.g., in this case, $\texttt{i64\ } 21$. At this point, it is possible to replace $\texttt{i64\ } 21$ with the use of $\%0$ and remove the original \texttt{call} to the $@array\_get\_size\_1d$ function, which is finally shown as the resulting code in Listing~\ref{lst:afterul}.

\begin{figure}[htb]
\centering
\begin{lstlisting}[caption=QIR code before \textbf{QIR-Loop-Unroll} optimization, basicstyle=\scriptsize\ttfamily,  numbers=left, style=mystyle, 
%linerange={22,28-30,34,119-131}, 
%xleftmargin=-0.5em, 
%xrightmargin=-3em, 
label={lst:beforeul}, 
breaklines=true]
br_1: 
  %qubits.i = tail call %Array* @qubit_allocate_array(~i64 21~)
  ^%0 = tail call i64 @array_get_size_1d(%Array* %qubits.i)^
  ^%1 = add i64 %0, -1^
  %.not = icmp slt i64 %1, 0
  br i1 %.not, label %br_2, label %br_1

\end{lstlisting}

\begin{lstlisting}[caption=QIR code after the \textbf{QIR-Loop-Unroll} optimization, basicstyle=\scriptsize\ttfamily,  numbers=left, style=mystyle, 
%linerange={22,28-30,34,119-131}, 
%xleftmargin=-0.5em, 
%xrightmargin=-3em, 
label={lst:afterul}, 
breaklines=true]
br_1: 
  %qubits.i = tail call %Array* @qubit_allocate_array(~i64 21~)
  ^%0 = add^ ~i64 21~^, -1^
  %.not = icmp slt i64 %0, 0
  br i1 %.not, label %br_2, label %br_1

\end{lstlisting}

    \caption{An example after optimization using \textbf{QIR-Loop-Unroll}. In this case, we replace values only available through QIR-specific functions with constant values, thus providing the opportunity for loop unrolling.}
    \centering
\end{figure}

\subsubsection{QDFO Execution}
\label{subsec:QDFO}

This subsection presents two qubits dataflow-based optimization methods for QIR, \textbf{QDFO-Load} and \textbf{QDFO-Create}, which reduce the repetitive $Load\_Op$ and $Create\_Op$ in the QIR code, respectively. To eliminate the QIR-specific dead code left after LLVM O3 optimization, we additionally designed \textbf{QIR-DCE}. Meanwhile, for the memory management problem caused by the optimization of \textbf{QDFO-Create}, we designed an optimization method for memory management and built it into \textbf{QDFO-Create}. The \textbf{QIR-CtlInline} method in the workflow will not be described separately because it only searches the QIR code for functions of the controlled gates and the SWAP gates and adds the \texttt{always-inline} attributes.

\paragraph{QDFO-Load}

\begin{figure}[!htb]
    \centering
    \includegraphics[width=1\textwidth]{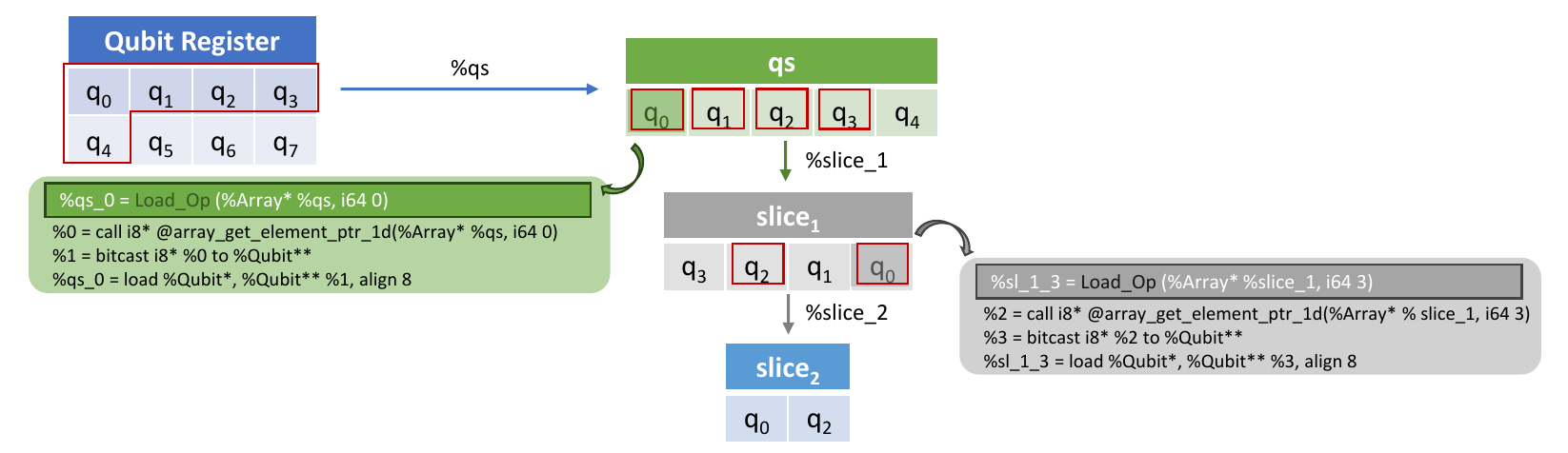}
    \caption{An example for qubit array allocation and slice operations.}
    \label{fig:LoadOp}
\end{figure}

The \textbf{QDFO-Load} method works to eliminate duplicate $Load\_Op$. As described in Section~\ref{sec:ME}, the reason for the resulting duplicate $Load\_Op$ is that the compiler is unable to obtain the pointer to the \texttt{\%Qubit} object directly from the QIR code, which can only be obtained through dynamic execution, and therefore has to call the function that obtains the pointer before each quantum gate operation is performed. We propose to analyze the QIR code statically to get as much information about the relationship between each \texttt{\%Qubit} object and the hardware registers in the code as possible. 

To better illustrate our approach, we first need to describe the source of the content loaded by $Load\_Op$. The execution of $Load\_Op$ presupposes the existence of a qubit array, which can be allocated from registers by the function $@qubit\_allocate\_array$ ($\%qs$ in Figure~\ref{fig:LoadOp}), or sliced from an existing qubit array by the function $@array\_slice\_1d$ ($\%slice_1$ and $\%slice_2$ in Figure~\ref{fig:LoadOp}). Since the slice operation is equivalent to mapping the original \texttt{\%Array} object, it is necessary to perform additional processing on the \texttt{\%Array} object obtained by the slice compared to the allocated qubit arrays directly. Specifically, $@array\_slice\_1d$ passes in three parameters \texttt{\%Array*} array, \texttt{\%Range} range and i1 true, where array denotes the parent \texttt{\%Array} object on which the slice operation is performed, \texttt{\%Range} is a QIR-specific type that can be used to represent an ordered array, and i1 true is used for the management of its alias. 

\begin{algorithm}
	\caption{$Slice\_Calculating(I, SliceInfo\_list) = $}
	\label{alg:slicecal}
        \begin{algorithmic}[1]
        \IF{$I$ is not a slice creating instruction}\RETURN
        \ENDIF
        \STATE \textcolor{teal}{// struct SliceInfo: }
        \STATE \textcolor{teal}{// \quad\quad string tag}
        \STATE \textcolor{teal}{// \quad\quad Instruction* SliceI}
        \STATE \textcolor{teal}{// \quad\quad Instruction* SliceFromI}
        \STATE \textcolor{teal}{// \quad\quad vector<int> qRef}
        \STATE \textcolor{teal}{// Prepare a SliceInfo object to store the slice information for instruction I}
        \STATE $SliceInfo\ sliceElm$ 
        \STATE $sliceElm.SliceI=I$ \textcolor{teal}{\quad// Stores I to represent the slice}
        \STATE \textcolor{teal}{// Get the source array and index from I}
        \STATE $Parent\_Array=I.getOperand(0)$
        \STATE $Range=I.getOperand(1)$
        \STATE $Index\_list=range\_calculating(Range)$
        \IF{$Parent\_Array$ is a slice creating instruction}
        \STATE $sliceElm.tag="slice"$
        \FOR{$temp\_sliceElm$ in $SliceInfo\_list$}
        \IF{$temp\_sliceElm.SilceI == Parent\_Array$}
        \STATE \textcolor{teal}{// If it comes from a slice, then they belong to the same allocated qubit array}
        \STATE $sliceElm.SliceFromI=temp\_sliceElm.SliceFromI$
        \STATE \textcolor{teal}{// The indices are calculated from the qRef of the parent's slice}
        \FOR{$i$ in $Index\_list$}
        \STATE $sliceElm.qRef.append(temp\_sliceElm.qRef[i])$
        \ENDFOR
        \STATE break
        \ENDIF
        \ENDFOR
        \ELSIF{$Parent\_Array$ is a qubit array allocating instruction}
        \STATE $sliceElm.tag="array"$
        \STATE $sliceElm.SliceFromI=Parent\_Array$
        \STATE $sliceElm.qRef=Index\_list$
        \ELSE 
        \RETURN
        \ENDIF
        \STATE \textcolor{teal}{// Store the information of the slice generated by instruction I into SliceInfo\_list}
        \STATE $SliceInfo\_list.append(sliceElm)$
        \RETURN
        \end{algorithmic}
\end{algorithm}

We summarize the processing for the slicing instruction in the Algorithm~\ref{alg:slicecal}. The main task of this algorithm is to iterate through the instructions in the function of the QIR code and collect the information of all the creating slice instructions in the list $SliceInfo\_list$ for subsequent use. In line 10, we declare a variable $sliceElm$ of type \texttt{SliceInfo} to store information about the slice. SliceInfo is a self-defined structure, which contains four attributes (lines 4-8): \texttt{tag} indicates the source of the slice, which is assigned as "array" when it comes from an allocated qubit array, and "slice" when it comes from another slice; \texttt{SliceI} stores the instruction that created the slice; \texttt{SliceFromI} stores the allocated qubit array from which the slice originate; together with the attribute \texttt{qRef}, the qubit corresponding to the original allocated qubit array can be obtained directly for each qubit stored in the slice. After storing the instruction $I$ itself into $sliceElm$ (line 11), the algorithm extracts the arrays it originated from and the indices of the slice elements in the source arrays from $I$'s operands, respectively (lines 13-15). After this, the algorithm is processed separately based on the type of the source array (lines 16-35). As an example, in Figure~\ref{fig:LoadOp}, the source of $slice_1$ is allocated qubit array, and the source of $slice_2$ is slice, where the three instructions $\%qs$, $\%slice\_1$, $\%slice\_2$ are: 
\begin{list}{}{\setlength{\itemsep}{0em}}
\small
    \item $\%qs = call\ \%Array*\ @qubit\_array\_allocate(i64\ 5)$
    \item $\%slice\_1 = call\ \%Array*\ @array\_slice\_1d(\%Array*\ \%qs, \{i64\ 3, i64\ -1, i64\ 0\}, i1\ true)$
    \item $\%slice\_2 = call\ \%Array*\ @array\_slice\_1d(\%Array*\ \%slice\_1, \{i64\ 3, i64\ -2, i64\ 0\}, i1\ true)$
\end{list}
From the second operand of $slice_1$ we can calculate that the index is \{3, 2, 1, 0\}, so in $SliceInfo\_list$, according to line 32, its \texttt{qRef} should be: 
\begin{equation}
    slice\_1.qRef = (3, 2, 1, 0). 
\end{equation}
Similarly, we can compute the index of $slice_2$ as \{3, 1\}, according to line 23-24, its \texttt{qRef} should be: 
\begin{equation}
    slice\_2.qRef = (slice\_1.qRef[3], slice\_1.qRef[1]) = (0, 2). 
\end{equation}
According to Algorithm~\ref{alg:slicecal}, the \texttt{SliceFromI} of these two slice-creating instructions should be the same allocated qubit array (line 21 and line 31). If the source array is not from either of the above two cases, it will not be saved, and the processing of the $sliceElm$ is interrupted (line 34). Finally, the $sliceElm$ is stored in $SliceInfo\_list$ (line 37), and the algorithm's execution ends. 

Since loading \texttt{\%Qubit*} objects from \texttt{\%Array} has a fixed sequence of instructions (e.g., $\%qs\_0$ and $\%sl\_1\_3$ in Figure~\ref{fig:LoadOp}), it is possible to start with the calling instruction of the $@array\_get\_element\_ptr\_1d$ function, that is, the first instruction of $Load\_Op$, and search and check it by using the Use-Define chain in the LLVM, to obtain all $Load\_Op$'s in the function. We take an abbreviated notation for $Load\_Op$, such as $\%qs\_0$ in Figure~\ref{fig:LoadOp}, which we notate as $\%qs\_0 = Load\_Op(\%Array*\ \%qs, i64\ 0)$. The $\%qs\_0$ is the instruction "$\%qs\_0 = load\ \%Qubit*,\ \%Qubit**\ \%1,\ align\ 8$", $\%Array*\ \%qs$ and $i64\ 0$ are the first and second operands of the $@array\_get\_element\_ptr\_1d$ function. 

We use $LoadOp\_list$ to collect all the $Load\_Op$ in the function and together with $SliceInfo\_list$ as inputs to the function shown in Algorithm~\ref{alg:QDFOLoad}. To reduce the number of loops and if statements in the pseudo-code, we have simplified the algorithm, mainly in the sense that in this algorithm, we assume that the $Load\_Op$'s in the $LoadOp\_list$ all come from the same allocated qubit array. In the actual algorithm used, we utilize the \texttt{SliceFromI} attribute in \texttt{SliceInfo} and the first operand of the function $@array\_get\_element\_ptr\_1d$ to identify the allocated qubit array from which it originated.

\begin{algorithm}
	\caption{: \textbf{QDFO-Load}(LoadOp\_list, SliceInfo\_list) = }
	\label{alg:QDFOLoad}
        \begin{algorithmic}[1]
        \STATE \textcolor{teal}{// Assume that the max qubit number is 64}
        \STATE $qubitTable = [64][]$
        \STATE \textcolor{teal}{// Load\_Op sample: $\%qs\_0 = Load\_Op(\%Array*\ \%qs, i64\ 0)$}
        \FOR{$temp\_LoadOp$ in $LoadOp\_list$}
        \STATE \textcolor{teal}{// 1st operand: $\%Array*\ \%qs$}
        \STATE $Source\_Array = temp\_LoadOp.getOperand(0)$
        \STATE \textcolor{teal}{// 2nd operand: $i64\ 0$}
        \STATE $indice = temp\_LoadOp.getOperand(1)$
        \IF{$Source\_Array$ is a slice creating instruction}
        \FOR{$sliceElm$ in $SliceInfo\_list$}
        \IF{$sliceElm.SliceI==Source\_Array$}
        \STATE \textcolor{teal}{// $sliceElm.qRef[indice]$ is the corresponding index in allocated qubit array}
        %\STATE \textcolor{teal}{// Store $temp\_LoadOp$ to the appropriate location in the allocated qubit array}
        \STATE $qubitTable[sliceElm.qRef[indice]].append(temp\_LoadOp)$
        \STATE break
        \ENDIF
        \ENDFOR
        \ELSIF{$Source\_Array$ is a qubit array allocating instruction}
        \STATE $qubitTable[indice].append(temp\_LoadOp)$
        \ENDIF
        \ENDFOR
        \FOR{$qubitTable[i]$ in $qubitTable$}
        \IF{$qubitTable[i].size$ > $1$}
        \FOR{different $qubitTable[i][j]$ and $qubitTable[i][k]$ in $qubitTable[i]$}
        \STATE \textcolor{teal}{// Determine which instruction comes first}
        \STATE $(beforeLoadOp, afterLoadOp)=isBefore(qubitTable[i][j], qubitTable[i][k])$
        \STATE \textcolor{teal}{// Replace all uses of afterLoadOp in the code with beforeLoadOp}
        \STATE \textcolor{teal}{// $I.getParent().getParent()$ returns the function where instruction $I$ located}
        \FORALL{$I$ in $qubitTable[i][j].getParent().getParent()$}
        \FORALL{$Use$ in $I.operand()$}
        \IF{$Use.get()==afterLoadOp$}
        \STATE $Use.set(beforeLoadOp)$
        \ENDIF
        \ENDFOR
        \ENDFOR
        \ENDFOR
        \ENDIF
        \ENDFOR
        \end{algorithmic}
\end{algorithm}

The core idea of our \textbf{QDFO-Load} algorithm is to try to locate all the $Load\_Op$'s in the function on the allocated qubit array so that we can exclude duplicate $Load\_Op$'s using the qubits of the allocated qubit array as a reference. So in Algorithm~\ref{alg:QDFOLoad}, we first create a vector $qubitTable$ to categorize the $Load\_Op$ for each qubit (line 2). In lines 4-20, we categorized the elements in $LoadOp\_list$. Specifically, we first distinguish where the source of the load operation is: if the source is a slice, the corresponding \texttt{SliceInfo} object can be found with the help of the information stored in $SliceInfo\_list$, and the index of the loaded qubit on the allocated qubit array can be calculated from \texttt{qRef} (lines 9-16); if the source is an allocated qubit array, the $indice$ is exactly the index (lines 17-19).
At the end of the traversal of the $LoadOp\_list$, the $qubitTable[i]$ stores all the $Load\_Op$ that is equivalent to loaded $i^{th}$ qubit of the allocated qubit array (line 20). After identifying the earliest $Load\_Op$ in $qubitTable[i]$, we can utilize LLVM's Define-Use chain to locate all instances of the duplicate $Load\_Op$. These instances will be replaced with the earliest $Load\_Op$ to complete the optimization (lines 21-36). 

After executing this algorithm, it needs to perform the O3-level optimization method of LLVM and the \textbf{QIR-DCE} algorithm we designed once to complete the dead code elimination so that the specific optimization effect will be shown in Section~\ref{par:QDCE}.

\begin{figure}[htb]
\centering
\begin{lstlisting}[caption=QIR code of the first H gate and the first CNOT gate in Figure 2, basicstyle=\scriptsize\ttfamily,  numbers=left, style=mystyle, 
%linerange={22,28-30,34,119-131}, 
%xleftmargin=-0.5em, 
%xrightmargin=-3em, 
label=beforeQDFOLoad,
breaklines=true]
  |%qs = tail call %Array* @qubit_allocate_array(i64 3)|
  %0 = tail call i8* @array_get_element_ptr_1d(~%Array* %qs~, ~i64 2~)
  %1 = bitcast i8* %0 to %Qubit**
  ^%qubit = load %Qubit*, %Qubit** %1, align 8^
  tail call void @qis__h__body(%Qubit* %qubit)
  ...
  %5 = tail call i8* @array_get_element_ptr_1d(~%Array* %qs~, ~i64 2~)
  %6 = bitcast i8* %5 to %Qubit**
  ^%7 = load %Qubit*, %Qubit** %6, align 8^
  |%__controlQubits__.i = tail call %Array* @array_create_1d(i32 8, i64 1)|
  |%8 = tail call i8* @array_get_element_ptr_1d(%Array* %__controlQubits__.i, i64 0)|
  |%9 = bitcast i8* %8 to %Qubit**|
  |store %Qubit* %4, %Qubit** %9, align 8|
  tail call void @qis__x__ctl(%Array* %__controlQubits__.i, ~%Qubit* %7~)

\end{lstlisting}

\begin{lstlisting}[caption=Optimized QIR code of the first H gate and the first CNOT gate in Figure 2, basicstyle=\scriptsize\ttfamily,  numbers=left, style=mystyle, 
%linerange={22,28-30,34,119-131}, 
%xleftmargin=-0.5em, 
%xrightmargin=-3em, 
label=afterQDFOLoad, 
breaklines=true]
  |%qs = tail call %Array* @qubit_allocate_array(i64 3)|
  %0 = tail call i8* @array_get_element_ptr_1d(~%Array* %qs~, ~i64 2~)
  %1 = bitcast i8* %0 to %Qubit**
  ^%qubit = load %Qubit*, %Qubit** %1, align 8^
  tail call void @qis__h__body(%Qubit* %qubit)
  ...
  |%__controlQubits__.i = tail call %Array* @array_create_1d(i32 8, i64 1)|
  |%8 = tail call i8* @array_get_element_ptr_1d(%Array* %__controlQubits__.i, i64 0)|
  |%9 = bitcast i8* %8 to %Qubit**|
  |store %Qubit* %4, %Qubit** %9, align 8|
  tail call void @qis__x__ctl(%Array* %__controlQubits__.i, ^%Qubit* %qubit^)

\end{lstlisting}
    \caption{An example of an optimization process for $Load\_Op$. Since $\%qubit$ and $\%7$ point to the same qubit, $\%7$ is optimized.}
    \centering
    \label{fig:loadcode}
\end{figure}

\paragraph{QIR-DCE}
\label{par:QDCE}
%\textcolor{red}{DCE for the QIR. LLVM does not do the DCE on the qir code because of the side-effect of the QIR instructions. }

\textbf{QIR-DCE} mainly implements dead code elimination of QIR-specific function calls in the QIR code, e.g., pointers and slice arrays that are no longer used during the optimization of \textbf{QDFO-Load}. The QIR code for the first H gate and the first CNOT gate in Figure~\ref{fig:toffoligate} is shown in Figure~\ref{fig:loadcode}: Since $\%qubit$ and $\%7$ refer to the same qubit, after processing by the \textbf{QDFO-Load} function, the second operand of the code in line 10 of Listing~\ref{beforeQDFOLoad} calling $@qis\_\_x\_\_ctl$ in the instruction will be replaced with $\%Qubit* \%qubit$. Then, the instructions $\%8$ and $\%9$ will become dead code and be deleted when the LLVM optimizer performs O3-level optimization. Although $\%5$ is also dead code, since it is a \texttt{call} instruction, the LLVM optimizer cannot determine if there is a side effect on the function called and, therefore, will not perform DCE. In \textbf{QIR-DCE}, we start by defining a list of keywords that include the names of the functions we want to check, such as "get\_element\_ptr" and "array\_slice." After that, \textbf{QIR-DCE} traverses all \texttt{call} instructions $I$ in function $F$ that contain a list of keywords and obtains all uses of instruction $I$ with the Define-Use chain in LLVM. If the fetched uses are null or only memory management-related function \texttt{call} instructions are present, these uses will be eliminated as dead code along with instruction $I$. Listing~\ref{afterQDFOLoad} shows the result of the final optimization. 

\begin{comment}

\begin{algorithm}
	\caption{: \textbf{QIR-DCE}(Function \&F) = }
	\label{alg:QIRDCE}
        \begin{algorithmic}[1]
        \STATE $keyword\_list = ["slice", "get\_ptr"]$
        \STATE $deleteI\_list = []$
        \FOR{$BasicBlock\ \&BB$ in $F$}
        \FOR{$Instruction\ \&I$ in $BB$}
        \IF{$I$ is not dependented}
        \IF{$I$ is a $CallInst$} \IF{$I.called\_function\_name().contains (elm$ in $keyword\_list)$}
        \STATE $deleteI\_list.append(I)$
        \ENDIF
        \ENDIF
        \ENDIF
        \ENDFOR
        \ENDFOR
        \FOR{$I$ in $deleteI\_list$}
        \STATE $I.ereaseFromParent()$
        \ENDFOR    
        \end{algorithmic}
\end{algorithm}
\end{comment}

\paragraph{QDFO-Create}
\label{par:QDFOCreate}

\begin{figure}[!htb]
    \centering
    \includegraphics[width=1\textwidth]{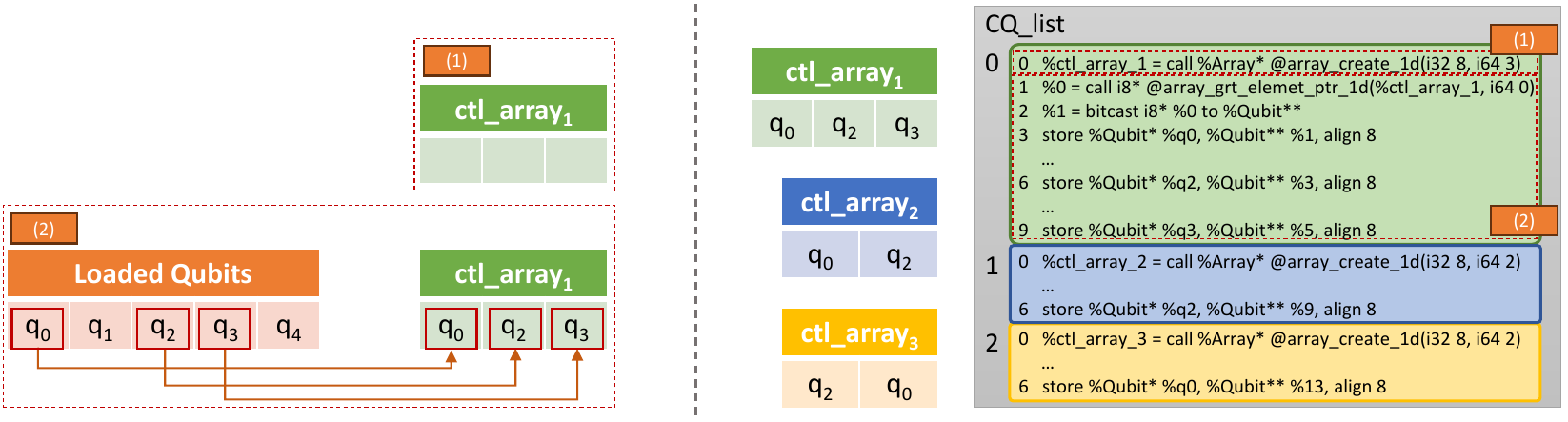}
    \caption{An example about $Create\_Op$. The figure on the left shows the two steps of $Create\_Op$, and the figure on the right shows how the created control qubit array is stored in the $CQ\_list$.}
    \label{fig:CreateOp}
\end{figure}

\textbf{QDFO-Create} is designed to eliminate repetitive qubit arrays creating operations. To fulfill this requirement, we first collect all $Create\_Op$ in the function. Figure~\ref{fig:CreateOp} shows the $Create\_Op$ process. Specifically, a $Create\_Op$ consists of the following steps: (1) calling the function $@array\_create\_1d$ to create an \texttt{\%Array} object with a size that controls the number of qubits, and (2) storing the \texttt{\%Qubit*} object to the appropriate index in the control qubit array. With the Define-Use chain in LLVM, all the instructions of the corresponding $Create\_Op$ can be obtained from each $@array\_create\_1d$ function \texttt{call} instruction and stored in a two-dimensional vector $CQ\_list$. For each element $CQElm$ in the $CQ\_list$, stored in $CQElm[0]$ is the instruction that calls creating that control qubit array, which will be utilized as a control gate operation. 
After this, the instructions for storing loaded \texttt{\%Qubit*} objects into \texttt{\%Array} are stored in groups of three instructions, including a \texttt{call} instruction to $@array\_get\_element\_ptr\_1d$ to obtain a pointer to the specified location from a control qubit array, a \texttt{bitcast} instruction to convert the pointer to type \texttt{\%Qubit**} (which is also used to determine whether $CQElem[0]$ is a control qubit array), and a \texttt{store} instruction that stores the address of the specified qubit register into the array. We show an example of a $CQ\_list$ in Figure~\ref{fig:CreateOp}.
%, where $CQ\_list[0]$ stores a $Create\_Op$ of $\%ctl\_array\_1$. $CQ\_list[0][0]$ stores the instructions for which it was created. From $CQ\_list[0][1]$ to $CQ\_list[0][3]$, $CQ\_list[0][4]$ to $CQ\_list[0][6]$, and $CQ\_list[0][7]$ to $CQ\_list[0][9]$ stores the \texttt{store} instructions for storing $\%q0$, $\%q2$, and $\%q3$ into $\%ctl\_array\_1$, respectively.
Since the \texttt{store} instruction in which the \texttt{\%Qubit*} object is stored into the \texttt{\%Array*} object does not have a return value, even if the control qubit array generated by $Create\_Op$ is not used, the instructions contained therein will not be recognized as dead code by the compiler. So unlike \textbf{QDFO-Load}, the elimination of duplicate $Create\_Op$ will be done during the execution of \textbf{QDFO-Create}.

\begin{algorithm}
	\caption{: \textbf{QDFO-Create}(CQ\_list) = }
	\label{alg:QDFOStore}
        \begin{algorithmic}[1]
        \STATE $wait4deleteI = []$
        \FOR{different index $i$ and $j$ in $CQ\_list$}
        \IF{$CQ\_list[i].size == CQ\_list[j].size$}
        \STATE \textcolor{teal}{// The 2nd operand of $@array\_create\_1d$ is the length of the creating array}
        \STATE $length = CQ\_list[i][0].getOperand(1)$
        \STATE $CQ\_i\_Elm\_set = []$
        \STATE $CQ\_j\_Elm\_set = []$
        \FOR{$indice\gets 0$ to $length$}
        \STATE \textcolor{teal}{// $CQ\_list[i][3*indice+3].getOperand(0)$: Get the stored \texttt{\%Qubit*} object.}
        \STATE $CQ\_i\_Elm\_set.insert(CQ\_list[i][3*indice+3].getOperand(0))$
        \STATE $CQ\_j\_Elm\_set.insert(CQ\_list[j][3*indice+3].getOperand(0))$
        \ENDFOR
        \IF{$CQ\_i\_Elm\_set==CQ\_j\_Elm\_set$}
        \STATE \textcolor{teal}{// Determine which command comes first}
        \STATE $(beforeCreateI, afterCreateI)=isBefore(CQ\_list[i][0], CQ\_list[j][0])$
        \STATE \textcolor{teal}{// Replace all uses of afterCreateI in the code with beforeCreateI}
        \STATE \textcolor{teal}{// $I.getParent().getParent()$ returns the function where instruction $I$ located}
        \FORALL{$I$ in $CQ\_list[i][0].getParent().getParent()$}
        \FORALL{$Use$ in $I.operand()$}
        \IF{$Use.get()==afterCreateI$}
        \STATE $Use.set(beforeCreateI)$
        \ENDIF
        \ENDFOR
        \ENDFOR
        \IF{$beforeCreateI == CQ\_list[i][0]$}
        \FORALL{$I$ in $CQ\_list[j]$}
        \STATE $wait4deleteI.append(I)$
        \ENDFOR
        \ELSE 
        \FORALL{$I$ in $CQ\_list[i]$}
        \STATE $wait4deleteI.append(I)$
        \ENDFOR
        \ENDIF
        \ENDIF
        \ENDIF
        \ENDFOR
        \STATE \textcolor{teal}{// Remove the collected repetitive Create\_Op}
        \FORALL{$I$ in $wait4deleteI$}
        \STATE $I.eraseFromParent()$
        \ENDFOR
        \end{algorithmic}
\end{algorithm}

Algorithm~\ref{alg:QDFOStore} illustrates the specific process of \textbf{QDFO-Create}. At the beginning of the algorithm, we first prepare a list $wait4deleteI$ to store the instructions to be deleted (line 1). Then, for any two same-sized elements $CQ\_list[i]$ and $CQ\_list[j]$ in $CQ\_list$, we obtain the qubit register addresses stored in them respectively to compare whether they store the same qubits (lines 3-12). A set is chosen to store these qubit register addresses because the order of the qubits in the control qubit array does not affect the effect of the CNOT gate operation (lines 6-7). If they store the same qubits, the instructions that created the arrays in $CQ\_list[i]$ and $CQ\_list[j]$ are compared. The instruction created earlier replaces all later-created instructions in the function with the one created earlier. Then, the elements of the $CQ\_list$ to which the later-created instruction belongs are added to $wait4deleteI$ (lines 13-34). Finally, after all the elements in the $CQ\_list$ have been traversed, all the instructions in $wait4deleteI$ are deleted (lines 38-39), thus deleting all duplicate $Create\_Op$. 

The QIR code for Figure~\ref{fig:storecode} comes from the last 3 CNOT gates in Figure~\ref{fig:toffoligate}. By analyzing lines 2-16 in Listing~\ref{beforeQDFOStore} we can find that the three arrays in the code $\%\_\_controlQubits\_\_.i$, $\%\_\_controlQubits\_\_.i1$, $\%\_\_controlQubits\_\_.i2$ are the same control qubit arrays, which all store $\%Qubit*\ \%q$ and belong to the $Create\_Op$ that is a duplicate. After \textbf{QDFO-Create} processing, the code in Listing~\ref{afterQDFOStore} can be obtained, as a result of optimization, the first created control qubit array $\%\_\_controlQubits\_\_.i$ is retained, the parameters of the CNOT gate operation in line 5, line 7, and line 9 are correctly replaced, and the duplicate $Create\_Op$ was deleted.

\begin{comment}
\begin{algorithm}
	\caption{: Controlled\_Qubit\_Arrays\_Calculating(Function \&F) = }
	\label{alg:CQAcal}
        \begin{algorithmic}[1]
        \STATE $CQ\_list = []$
        \FOR{$BasicBlock\ \&BB$ in $F$}
        \FOR{$Instruction\ \&I$ in $BB$}
        \STATE $???$
        \ENDFOR
        \ENDFOR            
        \RETURN $CQ\_list$
        \end{algorithmic}
\end{algorithm}
\end{comment}

\begin{figure}[htb]
\centering
\begin{lstlisting}[caption=QIR code of the last three CNOT gate in Figure 2, basicstyle=\scriptsize\ttfamily,  numbers=left, style=mystyle, 
%linerange={22,28-30,34,119-131}, 
%xleftmargin=-0.5em, 
xrightmargin=-2em, 
label=beforeQDFOStore,
breaklines=true]
  %__controlQubits__.i = tail call %Array* @array_create_1d(i32 8, i64 1)
  %0 = tail call i8* @array_get_element_ptr_1d(%Array* %__controlQubits__.i, i64 0)
  %1 = bitcast i8* %0 to %Qubit**
  store ~%Qubit* %q~, %Qubit** %1, align 8
  ^tail call void @qis__x__ctl(~%Array* %__controlQubits__.i~, %Qubit* %qubit)^
  ...
  %__controlQubits__.i1 = tail call %Array* @array_create_1d(i32 8, i64 1)
  %2 = tail call i8* @array_get_element_ptr_1d(%Array* %__controlQubits__.i1, i64 0)
  %3 = bitcast i8* %2 to %Qubit**
  store ~%Qubit* %q~, %Qubit** %3, align 8
  ^tail call void @qis__x__ctl(~%Array* %__controlQubits__.i1~, %Qubit* %qubit2)^
  ...
  %__controlQubits__.i2 = tail call %Array* @array_create_1d(i32 8, i64 1)
  %4 = tail call i8* @array_get_element_ptr_1d(%Array* %__controlQubits__.i2, i64 0)
  %5 = bitcast i8* %4 to %Qubit**
  store ~%Qubit* %q~, %Qubit** %5, align 8
  ^tail call void @qis__x__ctl(~%Array* %__controlQubits__.i2~, %Qubit* %qubit2)^

\end{lstlisting}

\begin{lstlisting}[caption=Optimized QIR code of the last three CNOT gate in Figure 2, basicstyle=\scriptsize\ttfamily,  numbers=left, style=mystyle, 
%linerange={22,28-30,34,119-131}, 
%xleftmargin=-0.5em, 
xrightmargin=-2em, 
label=afterQDFOStore,
breaklines=true]
  %__controlQubits__.i = tail call %Array* @array_create_1d(i32 8, i64 1)
  %0 = tail call i8* @array_get_element_ptr_1d(%Array* %__controlQubits__.i, i64 0)
  %1 = bitcast i8* %0 to %Qubit**
  store ~%Qubit* %q~, %Qubit** %1, align 8
  ^tail call void @qis__x__ctl(~%Array* %__controlQubits__.i~, %Qubit* %qubit)^
  ...
  ^tail call void @qis__x__ctl(~%Array* %__controlQubits__.i~, %Qubit* %qubit2)^
  ...
  ^tail call void @qis__x__ctl(~%Array* %__controlQubits__.i~, %Qubit* %qubit2)^

\end{lstlisting}
    \caption{An example of the optimization for $Create\_Op$. The control qubit array in line 1, line 7, and line 13 are constructed with the same qubit, so the last two are optimized. }
    \centering
    \label{fig:storecode}
\end{figure}

\begin{figure}[htb]
\centering
\begin{lstlisting}[caption=The QIR code at the end of performing the QIR\_store optimization. The $\%\_\_controlQubits\_\_.i$ will be freed at the end of the line 4\, resulting in a program error, basicstyle=\scriptsize\ttfamily,  numbers=left, style=mystyle, 
%linerange={22,28-30,34,119-131}, 
%xleftmargin=-0.5em, 
xrightmargin=-1em, 
label=beforeMMO,
breaklines=true]
  ^tail call void @array_update_alias_count(%Array* %__controlQubits__.i, ~i32 1~)^
  tail call void @qis__x__ctl(%Array* %__controlQubits__.i, %Qubit* %qubit)
  ^tail call void @array_update_alias_count(%Array* %__controlQubits__.i, ~i32 -1~)^
  ^tail call void @array_update_reference_count(%Array* %__controlQubits__.i, ~i32 -1~)^
  ...
  ^tail call void @array_update_alias_count(%Array* %__controlQubits__.i, ~i32 1~)^
  tail call void @qis__x__ctl(%Array* %__controlQubits__.i, %Qubit* %qubit)
  ^tail call void @array_update_alias_count(%Array* %__controlQubits__.i, ~i32 -1~)^
  ^tail call void @array_update_reference_count(%Array* %__controlQubits__.i, ~i32 -1~)^

\end{lstlisting}

\begin{lstlisting}[caption=The QIR code optimized for memory management maintains the correct lifecycle of qubit arrays, basicstyle=\scriptsize\ttfamily,  numbers=left, style=mystyle, 
%linerange={22,28-30,34,119-131}, 
%xleftmargin=-0.5em, 
xrightmargin=-1em, 
label=afterMMO,
breaklines=true]
  ^tail call void @array_update_alias_count(%Array* %__controlQubits__.i, ~i32 1~)^
  tail call void @qis__x__ctl(%Array* %__controlQubits__.i, %Qubit* %qubit)
  ...
  tail call void @qis__x__ctl(%Array* %__controlQubits__.i, %Qubit* %qubit)
  ^tail call void @array_update_alias_count(%Array* %__controlQubits__.i, ~i32 -1~)^
  ^tail call void @array_update_reference_count(%Array* %__controlQubits__.i, ~i32 -1~)^

\end{lstlisting}
    \caption{Once we have completed the QIR\_store optimization, if we don't optimize its memory-related functions accordingly, it will result in qubit arrays being released early, leading to program errors.}
    \centering
    \label{fig:memorycode}
\end{figure}

\textbf{Memory Management Optimization (MMO) }is to tune the calls to the \texttt{\%Array}'s allocation and release functions. In QIR, the \texttt{\%Array} objects have two attributes, alias and reference, which can be counted to control the timing of the release of these two objects. Since our optimization method has only ever processed objects generated via $@array\_create\_1d$ when executing \textbf{QIR-Create}, the processing of the \textbf{MMO} optimization is also limited to the \texttt{\%Array} object generated by this instruction. 
In our algorithm, for each control qubit array generated by calling the function $@array\_create\_1d$, we search for the first and last use of the instruction where it's called by a control gate operation and eliminate all memory management instructions in between. Since these created control qubit arrays are directly released after calling the control gate operation and do not undergo any other operations (as shown in lines 1-4 of Listing~\ref{beforeMMO}), adopting this elimination strategy is safe. Figure~\ref{fig:memorycode} shows an example after \textbf{QDFO-Create} optimization. In Listing~\ref{beforeMMO}, line 4's \texttt{call} to function $@array\_update\_reference\_count$ causes $\%\_\_controlQubits\_\_.i$ to be released prematurely, resulting in a program error when line 7 performs a CNOT gate operation on it. By executing our MMO method, the memory management code (lines 3-6 in Listing~\ref{beforeMMO}) will be eliminated so that $\%\_\_controlQubits\_\_.i$ will be released after the second CNOT gate operation (Listing~\ref{afterMMO}).
%Algorithm~\ref{alg:MMO} presents our optimization method. \textcolor{red}{need some explains}

\begin{comment}
\begin{algorithm}
	\caption{: Memory\_Management\_Optimization(Function \&F, Instruction* I) = }
	\label{alg:MMO}
        \begin{algorithmic}[1]
        \FOR{$BasicBlock\ \&BB$ in $F$}
        \FOR{$Instruction\ \&I$ in $BB$}
        \STATE $???$
        \ENDFOR
        \ENDFOR            
        \end{algorithmic}
\end{algorithm}
\end{comment}

\section{Experiment and Experience}\label{sec:EE}

 To demonstrate the effectiveness of our optimization algorithm, we implemented our approach as a custom pass on the LLVM optimizer. The implementation of our optimization method includes the implementation of the main algorithms such as \textbf{QIR-Inline}, \textbf{QDFO-Load}, and \textbf{QDFO-Create}, as well as the implementation of sub-functions such as detecting the loading of a qubit in the QIR code and detecting the sequential ordering of any two QIR instructions in a function. 

We have optimized the procedural implementation of several common quantum algorithms using our optimization algorithm. We initially attempted to use runtime as a validation metric, but because the matrix operations of the quantum circuit take much longer than the rest of the program when the QIR program is run on a simulator, it is difficult to observe the difference in program runtime before and after the program has been optimized by our algorithm. For real quantum computers, we attempted to upload the QIR code compiled from \texttt{Q\#} programs to Microsoft's cloud computing platform \texttt{Azure} for testing, but it failed to execute. Upon checking the QIR code samples provided by Microsoft, we noticed that some functions used in the QIR code were not defined in Microsoft's QIR specification documents. Therefore, we suspect Azure's computing platform is using a modified version of the QIR specification, which caused the issue. For the above reasons, the effect of our optimization is difficult to quantify. Thus, we chose to observe the changes in these QIR programs before and after optimization to extract some experience in optimizing using our algorithm.

\textbf{Experiment setup. }Our evaluation focuses on the programmatic implementation of 4 common quantum algorithms. Since there are no other optimization methods for QIR programs, we only show the effectiveness of our algorithms on code optimized by the LLVM optimizer at level O3. The compiling and executing tasks ran on a Dell G15 5515 laptop with a 3.2GHz 8-core AMD Ryzen7 5800 and 16GB RAM. The optimization tasks ran on a MacBook Pro with a 2.42-3.5GHz 12-core Apple M2 Pro and 16GB RAM. 

\subsection{Evaluation}

To fully validate the effectiveness of our approach, we conducted additional experiments based on the IBM Quantum Challenge Dataset [1]. It is designed to rigorously evaluate the performance of quantum optimization and qubit routing algorithms. We evaluate the effectiveness of our optimization approach by measuring the reduction in the number of instructions on the dataset. 

\begin{figure}[ht]
    \centering
    \includegraphics[width=1\textwidth]{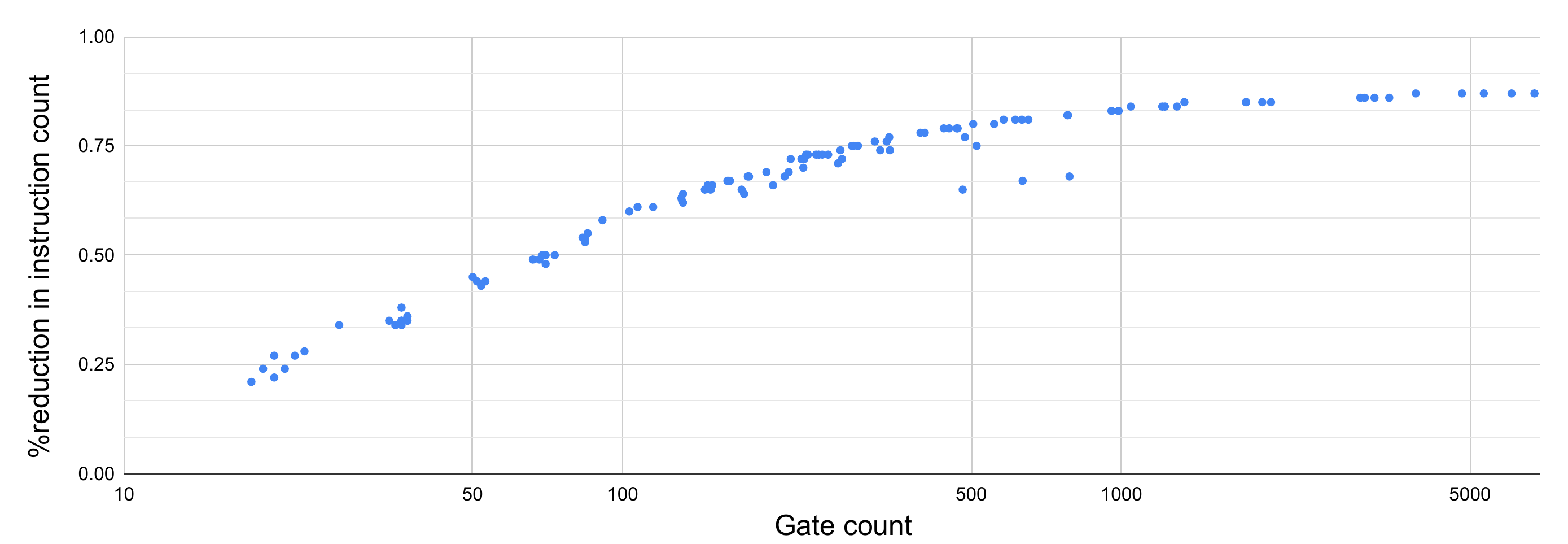}
    \caption{Experimental results on the IBM Quantum Challenge Dataset, where the horizontal coordinate is the number of quantum gates in the program and the vertical coordinate is the percentage of the number of instructions reduced by the optimization.}
    \label{fig:opt-reduction}
\end{figure}

Specifically, we selected a total of 120 quantum programs with quantum gate operations ranging from 18 to 6723 for our experiments, the results are shown in Figure~\ref{fig:opt-reduction}, where the horizontal coordinate indicates the number of quantum gates in each program, and the vertical coordinate indicates the ratio of the number of instructions reduced after optimization. The experimental results verify that our optimization method significantly reduces the number of instructions in the program for all different numbers of quantum gates, and the proportion of the reduced number of instructions becomes larger as the number of quantum gates increases (21\% reduction at 18 quantum gates and 87\% reduction at 6723 quantum gates). 

We believe that there are two main reasons for this result. The first is that the increased proportion of quantum computation in the program can provide more optimization opportunities for our method. The second is that as the number of quantum gates increases, the program generates redundant operations on the loading of qubits, and the generation of qubit arrays also increases. The experimental results verify the importance and effectiveness of our method, especially in larger quantum programs with better performance.

\subsection{Case Studies}

% In the course of our study, by comparing the QIR code before and after optimization, we found that in addition to the optimization of the number of instructions, there is also an optimization of the readability of the code. In this subsection, we will use two case studies to show the effect of our optimization algorithm on code readability. 
Our case studies focus on two aspects of discussion: whether our algorithms help link LLVM optimization methods to QIR code (\textbf{Connection}); and whether our algorithms reduce the number of duplicated $Load\_Op$ and $Create\_Op$ (\textbf{Optimization}).

\subsubsection{Grover's Algorithm}

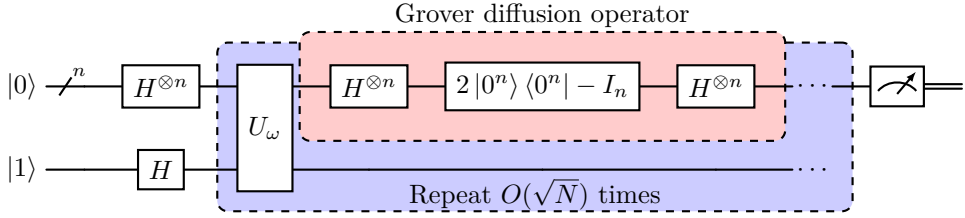
\begin{figure}[H]
    \centering
    \begin{quantikz}
    \ket{0}\
    &\qwbundle{n}&\gate{H^{\otimes n}}&\gate[2]{U_\omega}\gategroup[2,steps=5, style={dashed,rounded corners, fill=blue!20},background,label style={label
position=below}]{Repeat $O(\sqrt{N})$ times} &\gate{H^{\otimes n}}\gategroup[1,steps=3, style={dashed,rounded corners, inner sep=8pt, fill=red!20},background,label style={label
position=above}]{Grover diffusion operator}&\gate{2\ket{0^n}\bra{0^n}-I_n}&\gate{H^{\otimes n}}&\,\cdots\, & \meter{}&\setwiretype{c}\\
    \ket{1}\
    &&\gate{H}&&&&&\,\cdots\,
    \end{quantikz}
    \caption{Quantum circuit representation of Grover's algorithm}
    \label{fig:grover}
\end{figure}

Grover's algorithm\cite{grover1996fast} is a notable quantum algorithm renowned for its capacity to search for target items within an unsorted database. In contrast to conventional search algorithms such as linear or binary search, which typically exhibit linear or logarithmic time complexity, respectively, Grover's algorithm operates at a square root level, denoted as $O(\sqrt{N})$, where $N$ represents the number of items in the database. Quantum circuit representation of Grover’s algorithm is depicted in Figure \ref{fig:grover}. In this work, we use the \texttt{Q\#} Grover sample program~\cite{groversourcecode} provided by Microsoft as a source program to compile and optimize the resulting QIR code. 

\textbf{Connection: }For Grover's algorithm, our \textbf{QIR-Inline} algorithm optimizes 1 QIR-specific function called and computes the return value of the $@array\_get\_size\_1d$ function called at nine places within the program using the \textbf{QIR-Loop-Unroll} algorithm. By observing the results of the LLVM optimizer after applying O3-level optimization to the code, we confirmed that our algorithm contributes to the LLVM optimizer for loop unrolling and inlining for QIR code.

\textbf{Optimization:} We executed a complete optimization approach in Figure~\ref{fig:workflow} 2 times, which consumed a total of 0.284s. \textbf{QDFO-Load} and \textbf{QDFO-Create} reduced 83 and 8 lines of code, respectively, during the optimization. We cannot reduce the exact number of instructions because there are some classical parts of the computation in Grover's algorithm. The LLVM optimizer will optimize this part of the code during the execution of our optimization flow, which is mixed with the optimization of the quantum part, so it is difficult to directly count the specific reduction of the QIR part of the instructions. However, comparing the code before and after the optimization, we can observe that $Load\_Op$ is significantly reduced. As for $Create\_Op$, since Grover's algorithm has only a small number of CNOT gates, the optimization in this area is not obvious.

\subsection{Discussion}
In addition to the above two cases, we have also tried our optimization method on algorithms using different parameters or on some other common quantum algorithms such as Quantum Period Finding algorithm~\cite{shor1999polynomial, QPF}, which we can't enumerate for space reasons. Based on these practical attempts, we can obtain the following experience: 

\begin{itemize}
    \item In general, we found that our optimization methods can effectively preprocess QIR code, enabling LLVM optimization techniques to be applied to QIR code. Our algorithm allows some values that initially required execution to invoke QIR-specific functions to be statically analyzed during the compiler optimization. It provides more opportunities for the LLVM optimizer to perform optimizations. 
    \item Our optimization methods \textbf{QDFO-Load} and \textbf{QDFO-Create} can perform well on QIR code compiled by common quantum algorithmic program implementations. However, suppose a large amount of data in the program needs to be dynamically implemented before it can be obtained (e.g., the decision to allocate the length of a qubit array is made only at execution time). In that case, it will affect the analysis of the program by our algorithms, resulting in this part of the code not being optimized. In addition, QIR itself can perform the classical part of the operation; if the program uses the result of these QIR-specific operation functions to perform behaviors such as loop management, it will also cause our optimization to be less effective; this is because we have not designed the optimization algorithm for this part of the content.
    %\item Applying our algorithm effectively improves the readability of the code because our algorithm removes as much redundancy as possible from the QIR code and removes the different variables representing the same qubits and qubit arrays so that it is possible to identify exactly which objects each quantum gate operation acts on directly by the variable names. Such a representation is also closer to the high-level language representation. 
\end{itemize}

%\subsubsection{Shor's Algorithm}

\section{Related Work}\label{sec:RW}

\subsection{Quantum IRs}
\textcolor{black}{The intermediate representation is a data structure that most compilers translate into when faced with a source code program. Intermediate representations increase the possibilities for further analysis and processing by cleanly separating the front end from the back end. There have been numerous studies~\cite{grosser2012polly, necula2002cil, lattner2004llvm} in classical program compilers.}

\textcolor{black}{In recent years, as the potential of quantum programs has gained attention, numerous studies have focused on intermediate representations of quantum programs. McCaskey \emph{et al.}~\cite{mccaskey2021mlir} leverage the MLIR framework for quantum computing and extend it with a new quantum dialect for quantum compilation. Hietala \emph{et al.}~\cite{hietala2021verified} proposed a small quantum intermediate representation {SQIR}, which is a simple circuit-oriented language deeply embedded in the Coq platform~\cite{Coq-refman}. Microsoft also proposed a new intermediate representation {QIR}~\cite{Azure_QDK} for quantum programs based on the LLVM and specifies rules for quantum constructs. Although QIR is widely used due to its language- and hardware-agnostic, various opaque types pose challenges for static optimization by compilers. Our work improves this situation through a series of optimization methods.}

\subsection{Quantum Program Optimization}
Quantum programs have attracted attention for their potential to break through the upper limits of classical programs. As the preciousness of quantum resources~\cite{chitambar2019quantum} and the decoherence property~\cite{schlosshauer2019quantum} of a qubit, the optimization of quantum programs becomes an urgent problem to be solved. Amy \emph{et al.}~\cite{amy2013meet} proposed an algorithm for finding the minimum depth quantum circuit to implement a given operation. Their work provides significant speedup for many critical, logical quantum operations. Tao \emph{et al.}~\cite{tao2022giallar} utilized rewrite rules for quantum circuits to reduce quantum gates during the verification for the Qiskit~\cite{qiskit} quantum compiler. These works focus only on optimizing quantum circuits and do not address other aspects of the optimization program, such as simplification of the number of instructions. They can serve as references for our work, and in future endeavors, we can also integrate quantum circuit optimization into our work.

Ittah \emph{et al.}~\cite{ittah2021enabling} proposed a multi-level intermediate representation QIRO to enable dataflow optimization for quantum programs. Their work encodes the dataflow directly in the IR to leverage dataflow analysis for optimizations of mixed quantum-classical programs. We consider this work complementary to ours since the two optimization methods act on different platforms, and the optimization of MLIR and QIR can be performed on one optimization chain.

To the best of our knowledge, optimizations that target QIR programs do not exist yet. We think it is because, in the QIR specification, almost all functions are implemented by the backend computing platform, and many compiler-opaque variables are present. These results in the compiler having little information available in the QIR code, leading to difficulties in optimization. Our optimization approach draws on the semantics of QIR and, through static analysis, captures some of the otherwise opaque information and makes equivalent substitutions to the code, thus providing opportunities for further optimization. 

\section{Concluding Remarks}
\label{sec:CR}

In this work, we propose and implement the Quantum Intermediate Representation Data Flow-based Optimization algorithm QDFO to optimize the QIR programs. To reuse existing LLVM optimization methods as much as possible in the optimization process, we design algorithms such as \textbf{QIR-Inline}, \textbf{QIR-Loop-Unroll}, \emph{etc.}, which allow function inlining and loop unrolling in the LLVM optimizer to be applied to the QIR code. We verified the effectiveness of the methods by studying and observing a series of real-world QIR code cases. 

Our work is a first attempt to optimize QIR, and some improvements can be made based on our approach. Since most quantum programs nowadays are hybrid classical-quantum programs, and QIR can also represent and process classical data (e.g., string and unlimited-precision integers), optimizing these instructions is also essential. Furthermore, although our work can be seen as preparation for optimizing quantum circuits on QIR, we have not designed the optimization of quantum circuits, which can effectively enhance the performance of quantum programs. Finally, it is possible to design the syntax and semantics of QIR utilizing formal verification, thus verifying the correctness of the optimization method and ensuring that it does not lead to changes in program functionality before and after the optimization. 

%%
%% Bibliography
%%

%% Please use bibtex, 

\bibliography{lipics-v2021-sample-article}

\end{document}